\documentstyle[twoside,epsf,psfig,times]{article}
\def\beqa{\begin{eqnarray}}
\def\eeqa{\end{eqnarray}}
\def\beq{\begin{equation}}
\def\eeq{\end{equation}}
\catcode`\@=11
\long\def\@makefntext#1{
\protect\noindent \hbox to 3.2pt {\hskip-.9pt  
$^{{\eightrm\@thefnmark}}$\hfil}#1\hfill}               %CAN BE USED 

\def\@makefnmark{\hbox to 0pt{$^{\@thefnmark}$\hss}}    %ORIGINAL 

\def\ps@myheadings{\let\@mkboth\@gobbletwo
\def\@oddhead{\hbox{}
\rightmark\hfil\eightrm\thepage}   
\def\@oddfoot{}\def\@evenhead{\eightrm\thepage\hfil
\leftmark\hbox{}}\def\@evenfoot{}
\def\sectionmark##1{}\def\subsectionmark##1{}}

\oddsidemargin=\evensidemargin
\newcounter{sectionc}\newcounter{subsectionc}\newcounter{subsubsectionc}
\renewcommand{\section}[1] {\vspace{12pt}\addtocounter{sectionc}{1} 
\setcounter{subsectionc}{0}\setcounter{subsubsectionc}{0}\noindent 
        {\tenbf\thesectionc. #1}\par\vspace{5pt}}
\renewcommand{\subsection}[1] {\vspace{12pt}\addtocounter{subsectionc}{1} 
\setcounter{subsubsectionc}{0}\noindent 
{\bf\thesectionc.\thesubsectionc. {\kern1pt \bfit #1}}\par\vspace{5pt}}
\renewcommand{\subsubsection}[1] {\vspace{12pt}\addtocounter{subsubsectionc}{1}
        \noindent{\tenrm\thesectionc.\thesubsectionc.\thesubsubsectionc.
        {\kern1pt \tenit #1}}\par\vspace{5pt}}
\newcommand{\nonumsection}[1] {\vspace{12pt}\noindent{\tenbf #1}
        \par\vspace{5pt}}

\newcounter{appendixc}
\newcounter{subappendixc}[appendixc]
\newcounter{subsubappendixc}[subappendixc]
\renewcommand{\thesubappendixc}{\Alph{appendixc}.\arabic{subappendixc}}
\renewcommand{\thesubsubappendixc}
        {\Alph{appendixc}.\arabic{subappendixc}.\arabic{subsubappendixc}}

\renewcommand{\appendix}[1] {\vspace{12pt}
        \refstepcounter{appendixc}
        \setcounter{figure}{0}
        \setcounter{table}{0}
        \setcounter{lemma}{0}
        \setcounter{theorem}{0}
        \setcounter{corollary}{0}
        \setcounter{definition}{0}
        \setcounter{equation}{0}
        \renewcommand{\thefigure}{\Alph{appendixc}.\arabic{figure}}
        \renewcommand{\thetable}{\Alph{appendixc}.\arabic{table}}
        \renewcommand{\theappendixc}{\Alph{appendixc}}
        \renewcommand{\thelemma}{\Alph{appendixc}.\arabic{lemma}}
        \renewcommand{\thetheorem}{\Alph{appendixc}.\arabic{theorem}}
        \renewcommand{\thedefinition}{\Alph{appendixc}.\arabic{definition}}
        \renewcommand{\thecorollary}{\Alph{appendixc}.\arabic{corollary}}
        \renewcommand{\theequation}{\Alph{appendixc}.\arabic{equation}}
        \noindent{\tenbf Appendix \theappendixc #1}\par\vspace{5pt}}
\newcommand{\subappendix}[1] {\vspace{12pt}
        \refstepcounter{subappendixc}
        \noindent{\bf Appendix \thesubappendixc. {\kern1pt \bfit #1}}
        \par\vspace{5pt}}
\newcommand{\subsubappendix}[1] {\vspace{12pt}
        \refstepcounter{subsubappendixc}
        \noindent{\rm Appendix \thesubsubappendixc. {\kern1pt \tenit #1}}
        \par\vspace{5pt}}

\newcommand{\textlineskip}{\baselineskip=13pt}
\newcommand{\smalllineskip}{\baselineskip=10pt}

\newcommand{\copyrightheading}[1]
        {\vspace*{-2.5cm}\smalllineskip{\flushleft
        {\footnotesize International Journal of Modern Physics D, #1}\\
        {\footnotesize \copyright\kern2pt World Scientific Publishing
         Company}\\
         }}

\newcommand{\publisher}[2]{{\begin{center}\footnotesize\smalllineskip 
        Received #1\\
        Revised #2
        \end{center}
        }}

\def\abstracts#1#2#3{{
        \centering{\begin{minipage}{4.5in}\footnotesize\baselineskip=10pt
        \parindent=0pt #1\par 
        \parindent=15pt #2\par
        \parindent=15pt #3
        \end{minipage}}\par}} 

\renewenvironment{thebibliography}[1]
        {\frenchspacing
         \ninerm\baselineskip=11pt
         \begin{list}{\arabic{enumi}.}
        {\usecounter{enumi}\setlength{\parsep}{0pt}     
         \setlength{\leftmargin 12.7pt}{\rightmargin 0pt}%FOR 1--9 ITEMS
         \setlength{\itemsep}{0pt} \settowidth
        {\labelwidth}{#1.}\sloppy}}{\end{list}}

\newcounter{itemlistc}
\newcounter{romanlistc}
\newcounter{alphlistc}
\newcounter{arabiclistc}

\newcommand{\fcaption}[1]{
        \refstepcounter{figure}
        \setbox\@tempboxa = \hbox{\footnotesize Fig.~\thefigure. #1}
        \ifdim \wd\@tempboxa > 5in
           {\begin{center}
        \parbox{5in}{\footnotesize\smalllineskip Fig.~\thefigure. #1}
            \end{center}}
        \else
             {\begin{center}
             {\footnotesize Fig.~\thefigure. #1}
              \end{center}}
        \fi}

\newcommand{\tcaption}[1]{
        \refstepcounter{table}
        \setbox\@tempboxa = \hbox{\footnotesize Table~\thetable. #1}
        \ifdim \wd\@tempboxa > 5in
           {\begin{center}
        \parbox{5in}{\footnotesize\smalllineskip Table~\thetable. #1}
            \end{center}}
        \else
             {\begin{center}
             {\footnotesize Table~\thetable. #1}
              \end{center}}
        \fi}

\def\@citex[#1]#2{\if@filesw\immediate\write\@auxout
        {\string\citation{#2}}\fi
\def\@citea{}\@cite{\@for\@citeb:=#2\do
        {\@citea\def\@citea{,}\@ifundefined
        {b@\@citeb}{{\bf ?}\@warning
        {Citation `\@citeb' on page \thepage \space undefined}}
        {\csname b@\@citeb\endcsname}}}{#1}}

\newif\if@cghi
\def\cite{\@cghitrue\@ifnextchar [{\@tempswatrue
        \@citex}{\@tempswafalse\@citex[]}}
\def\citelow{\@cghifalse\@ifnextchar [{\@tempswatrue
        \@citex}{\@tempswafalse\@citex[]}}
\def\@cite#1#2{{$\null^{#1}$\if@tempswa\typeout
        {IJCGA warning: optional citation argument 
        ignored: `#2'} \fi}}

\def\pmb#1{\setbox0=\hbox{#1}
        \kern-.025em\copy0\kern-\wd0
        \kern.05em\copy0\kern-\wd0
        \kern-.025em\raise.0433em\box0}

\def\fnt#1#2{\footnotetext{\kern-.3em
        {$^{\mbox{\scriptsize #1}}$}{#2}}}

\def\fpage#1{\begingroup
\voffset=.3in
\thispagestyle{empty}\begin{table}[b]\centerline{\footnotesize #1}
        \end{table}\endgroup}

\def\runninghead#1#2{\pagestyle{myheadings}
\markboth{{\protect\footnotesize\it{\quad #1}}\hfill}
{\hfill{\protect\footnotesize\it{#2\quad}}}}
\font\tenrm=cmr10
\font\tenit=cmti10 
\font\tenbf=cmbx10
\font\bfit=cmbxti10 at 10pt
\font\ninerm=cmr9

\font\eightrm=cmr8

\def\qed{\hbox{${\vcenter{\vbox{                  %HOLLOW SQUARE
   \hrule height 0.4pt\hbox{\vrule width 0.4pt height 6pt
   \kern5pt\vrule width 0.4pt}\hrule height 0.4pt}}}$}}

\begin{document}
\setlength{\textheight}{7.7truein}    %FOR 2ND PAGE ONWARDS

\runninghead{Bianchi Type I Anisotropic Magnetized Cosmologica 
            Models With Varying  $\Lambda$} 
{A. Pradhan and O. P. Pandey}

\normalsize\textlineskip
\thispagestyle{empty}
\setcounter{page}{1}

\copyrightheading{}             {Vol.~0, No.~0 (2003) 000--000}

\vspace*{0.88truein}

\fpage{1}

\centerline{\bf BIANCHI TYPE I ANISOTROPIC MAGNETIZED  COSMOLOGICAL}
\vspace*{0.035truein}
\centerline{\bf MODELS WITH VARYING $\Lambda$}
\vspace*{0.37truein}
%\centerline{}
%\vspace*{0.015truein}
%\centerline{}
%\baselineskip=10pt
%\centerline{}
%\vspace*{10pt}
\centerline{\footnotesize ANIRUDH PRADHAN\footnote{E-mail: acpradhan@yahoo.com,
pradhan@iucaa.ernet.in (Corresponding Author)}~ ~ and OM PRAKASH PANDEY}
\vspace*{0.015truein}  
\centerline{\footnotesize\it Department of Mathematics, Hindu Post-graduate College,}
\baselineskip=10pt
\centerline{\footnotesize\it Zamania, Ghazipur 232 331, India}
%\vspace*{10pt}
%\centerline{\footnotesize }
%\vspace*{0.015truein}
%\centerline{\footnotesize\it}
%\baselineskip=10pt
%\centerline{\footnotesize\it }
\baselineskip=10pt
%\centerline{\footnotesize }
\vspace*{0.225truein}
\publisher{(received date)}{(revised date)}
\vspace*{0.21truein}
\abstracts{Bianchi type I magnetized cosmological models in the presence 
of a bulk viscous fluid are investigated. The source of the magnetic field 
is due to an electric current produced along $x$-axis. The distribution 
consists of an electrically neutral viscous fluid with an infinite 
electrical conductivity. The coefficient of bulk viscosity is assumed to 
be a power function of mass density. The cosmological constant $\Lambda$ 
is found to be positive and is a decreasing function of time which is 
supported by results from recent supernovae observations. The behaviour 
of the models in presence and absence of magnetic field are also 
discussed.}{}{}
%\vspace*{10pt}
%\keywords{The contents of the keywords}

%\textlineskip                  %) USE THIS MEASUREMENT WHEN THERE IS
%\vspace*{12pt}                 %) NO SECTION HEADING

\vspace*{1pt}\textlineskip      %) USE THIS MEASUREMENT WHEN THERE IS
%\section{Introduction}
%\vspace*{-0.5pt}
%%%%%%%%%%%%%%%%%%%%%%%%%%%%%%%%%%%%%%%%%%%%%%%%%%%%%%%%%%%%%%%%%%%%%%%%%%
%%%%%%%%%%%%%%%%%%%%%%%%%%%%%%%   SECTION 1  %%%%%%%%%%%%%%%%%%%%%%%%%%%%%
\section{Introduction}
\vspace*{-0.5pt}
The occurrence of magnetic fields on galactic scale is well-established 
fact today, and their importance for a variety of astrophysical phenomena
is generally acknowledged as pointed out Zeldovich et al.\cite{ref1} Also 
Harrison \cite{ref2} has suggested that magnetic field could have a 
cosmological origin. As a natural consequences we should include magnetic 
fields in the energy-momentum tensor of the early universe.
The choice of anisotropic cosmological models in Einstein system of field 
equations leads to the cosmological models more general than Robertson-Walker 
model.\cite{ref3} The presence of primordial magnetic fields in the early 
stages of the evolution of the universe has been discussed by several authors.
\cite{ref4} $^-$\cite{ref13} Strong magnetic fields can be created due to 
adiabatic compression in clusters of galaxies. Large-scale magnetic fields 
give rise to anisotropies in the universe. The  anisotropic pressure created 
by the magnetic fields dominates the evolution of the shear anisotropy and it 
decays slower than if the pressure was isotropic.\cite{ref14,ref15} Such 
fields can be generated at the end of an inflationary epoch.\cite{ref16}$^-$
\cite{ref20} Anisotropic magnetic field models have significant contribution 
in the evolution of galaxies and stellar objects. Bali and Ali\cite{ref21} 
obtained a magnetized cylindrically symmetric universe with an 
electrically neutral perfect fluid as the source of matter. Several authors
\cite{ref22}$^-$\cite{ref26} have investigated Bianchi type I cosmological 
models with a magnetic field in different context.\\
\newline
\par
Most cosmological models assume that the matter in the universe can be described 
by `dust'(a pressure-less distribution) or at best a perfect fluid. Nevertheless,
there is good reason to believe that - at least at the early stages of the universe 
- viscous effects do play a role.\cite{ref27}$^-$\cite{ref29} For example, the 
existence of the bulk viscosity is equivalent to slow process of restoring 
equilibrium states.\cite{ref30} The observed physical phenomena such as the 
large entropy per baryon and remarkable degree of isotropy of the cosmic 
microwave background radiation suggest analysis of dissipative effects in cosmology.
Bulk viscosity is associated with the GUT phase transition and string creation. 
Thus, we should consider the presence of a material distribution other than a 
perfect fluid to have realistic cosmological models (see Gr\o n\cite{ref31} for 
a review on cosmological models with bulk viscosity). The effect of bulk viscosity 
on the cosmological evolution has been investigated by a number of authors in the 
framework of general theory of relativity.\cite{ref32}$^-$\cite{ref43}\\  
\newline
\par
Models with a relic cosmological constant $\Lambda$ have received considerable 
attention recently among researchers for various reasons 
(see Refs.{\cite{ref44}}$^-${\cite{ref48}} and references therein). Some of the 
recent discussions on the cosmological constant ``problem'' and on cosmology 
with a time-varying cosmological constant by Ratra and Peebles,\cite{ref49} 
Dolgov\cite{ref50}$^-$\cite{ref52} and Sahni and Starobinsky\cite{ref53}
point out that in the absence of any interaction with matter or radiation, the 
cosmological constant remains a ``constant'', however, in the presence of
interactions with matter or radiation, a solution of Einstein equations and the 
assumed equation of covariant conservation of stress-energy with a time-varying 
$\Lambda$ can be found. For these solutions, conservation of energy requires 
decrease in the energy density of the vacuum component to be compensated by a 
corresponding increase in the energy density of matter or radiation. Earlier 
researchers on this topic, are contained in Zeldovich,\cite{ref54} Bertolami,
\cite{ref56,ref57} Weinberg,\cite{ref45} and Carroll, Press and Turner.
\cite{ref55} Recent observations by Perlmutter {\it et al.}
\cite{ref58} and Riess {\it et al.} \cite{ref59} strongly favour a significant 
and positive $\Lambda$. Their finding arise from the study of more than $50$ 
type Ia supernovae with redshifts in the range $0.10 \leq z \leq 0.83$ and 
suggest Friedmann models with negative pressure matter such as a cosmological 
constant, domain walls or cosmic strings (Vilenkin, \cite{ref60} Garnavich 
{\it et al.}\cite{ref61}). Recently, Carmeli and Kuzmenko\cite{ref62} have shown 
that the cosmological relativity theory (Behar and Carmeli\cite{ref63})
predicts the value $\Lambda = 1.934\times 10^{-35} s^{-2}$ for the cosmological 
constant. This value of $\Lambda$ is in excellent agreement with the measurements 
recently obtained by the High-Z Supernova Team and Supernova Cosmological Project 
(Garnavich {\it et al.};\cite{ref61} Perlmutter {\it et al.};\cite{ref58} 
Riess {\it et al.};\cite{ref59} Schmidt {\it et al.}\cite{ref64}). The main 
conclusion of these works is that the expansion of the universe is accelerating. 
\newline
\par
Several ans$\ddot{a}$tz have been proposed in which the $\Lambda$ term decays 
with time (see Refs. Gasperini,\cite{ref65,ref66} Berman,\cite{ref67} 
Freese {\it et al.},\cite{ref48}  $\ddot{O}$zer and Taha,\cite{ref48} 
Peebles and Ratra,\cite{ref68} Chen and Hu,\cite{ref69} Abdussattar and 
Viswakarma,\cite{ref70} Gariel and Le Denmat,\cite{ref71} Pradhan 
{\it et al.}\cite{ref45,ref72}). Of the special interest is the ansatz 
$\Lambda \propto S^{-2}$ (where $S$ is the scale factor of the Robertson-Walker 
metric) by Chen and Wu,\cite{ref69} which has been considered/modified by 
several authors ( Abdel-Rahaman,\cite{ref73} Carvalho {\it et al.},\cite{ref48} 
Waga,\cite{ref74} Silveira and Waga,\cite{ref48} Pradhan and Pandey,\cite{ref75} 
Vishwakarma\cite{ref76}).
\newline
\par
Recently Bali and Gokhroo\cite{ref77} obtained a Bianchi type I anisotropic 
magnetized cosmological model for perfect fluid distribution. Motivated by 
the situations discussed above, in this paper, we shall focus upon the problem 
of establishing a formalism for studying the general relativistic evolution of 
magnetic homogeneities in presence of bulk viscous in an expanding universe. 
We do this by extending the work of Bali and Gokhroo \cite{ref77} by including 
an electrically neutral bulk viscous fluid as the source of matter in the 
energy-momentum tensor. This paper is organised as follows. The metric and the 
field equations are presented in Sec.$2$. The Sec.$3$ includes the solution of 
the field equations and physical and geometrical features of the models. In 
Sec.$4$, we obtain the bulk viscous cosmological solutions in the absence of 
the magnetic field. In Sec. $5$, we discuss our main results and summarize 
the conclusions.\\   
\newline
\par
%%%%%%%%%%%%%%%%%%%%%%%%%%%%%%%%%%%%%%%%%%%%%%%%%%%%%%%%%%%%%%%%%%%%%%%%
%%%%%%%%%%%%%%%%%%%%    SECTION 2   %%%%%%%%%%%%%%%%%%%%%%%%%%%%%%%%%%%
\section{Field  Equations}
 We consider the Bianchi type I metric in the form  
\begin{equation} 
\label{eq1}  
ds^{2} = -dt^{2} + A^{2}dx^{2} + B^{2}dy^{2} + C^{2}dz^{2},
\end{equation} 
where $A$, $B$ and $C$ are functions of $t$ only. The energy momentum tensor
in the presence of bulk stress has the form
\begin{equation} 
\label{eq2}
T^{j}_{i} = (\rho + \bar{p})v_{i}v^{j} + \bar{p}g^{j}_{i} + E^{j}_{i},
\end{equation} 
where $E^{j}_{i}$ is the electro-magnetic field given by Lichnerowicz\cite{ref78}
as 
\begin{equation} 
\label{eq3}
E^{j}_{i} = \bar{\mu}\left[|h|^{2}\left(v_{i}v^{j} + \frac{1}{2}g^{j}_{i}\right)
- h_{i}h^{j}\right]
\end{equation} 
and
\begin{equation} 
\label{eq4}
\bar{p} = p - \xi v^{i}_{;i}
\end{equation} 
Here $\rho$, $p$, $\bar{p}$ and $\xi$ are the energy density, isotropic pressure
, effective pressure, bulk viscous coefficient respectively and $v^{i}$ is the 
flow vector satisfying the relation
\begin{equation} 
\label{eq5}
g_{ij} v^{i}v^{j} = - 1
\end{equation} 
$\bar\mu$ is the magnetic permeability and $h_{i}$ the magnetic flux vector
defined by
\begin{equation} 
\label{eq6}
h_{i} = \frac{1}{\bar{\mu}}~~ ^*F_{ji}v^{j}
\end{equation} 
where $^*F_{ij}$ is the dual electro-magnetic field tensor defined by Synge
\cite{ref79} to be
\begin{equation} 
\label{eq7}
^*F_{ij} = \frac{\sqrt-g}{2}\epsilon_{ijkl} F^{kl}
\end{equation} 
$F_{ij}$ is the electro-magnetic field tensor and $\epsilon_{ijkl}$ is the
Levi-Civita tensor density. Here, the comoving coordinates are taken to be 
$v^{1}$ = $0$ = $v^{2}$ = $v^{3}$ and $v^{4}$ = $1$. We take the incident 
magnetic field to be in the direction of $x$-axis so that $h_{1} \ne 0$, 
$h_{2}$ =  $0$ = $h_{3}$ = $h_{4}$. Due to assumption of infinite
conductivity of the fluid, we get $F_{14} = 0 = F_{24} = F_{34}$. The only non-
vanishing component of $F_{ij}$ is $F_{23}$. The first set of Maxwell's equation
\begin{equation} 
\label{eq8}
F_{ij;k} + F_{jk;i} + F_{ki;j} = 0
\end{equation}  
leads to
\begin{equation} 
\label{eq9} 
F_{23} = I \mbox{(const.)}
\end{equation}  
where the semicolon represents a covariant differentiation. Hence
\begin{equation} 
\label{eq10} 
h_{1} = \frac{AI}{\bar{\mu}BC}
\end{equation}  
The Einstein's field equations 
\begin{equation} 
\label{eq11} 
R^{j}_{i} - \frac{1}{2} R g^{j}_{i} + \Lambda g^{j}_{i} = - 8\pi T^{j}_{i}, ~ ~ 
\mbox{(c = 1, G = 1 in gravitational unit)}
\end{equation}  
for the line element (1) has been set up as
\begin{equation} 
\label{eq12} 
8\pi\left(\bar{p} - \frac{I^{2}}{2\bar{\mu}B^{2}C^{2}}\right) = -\frac{B_{44}}{B}
- \frac{C_{44}}{C} - \frac{B_{4}C_{4}}{BC} - \Lambda
\end{equation}  
\begin{equation} 
\label{eq13} 
8\pi\left(\bar{p} + \frac{I^{2}}{2\bar{\mu}B^{2}C^{2}}\right) = -\frac{A_{44}}{A}
- \frac{C_{44}}{C} - \frac{A_{4}C_{4}}{AC} - \Lambda
\end{equation}  
\begin{equation} 
\label{eq14} 
8\pi\left(\bar{p} + \frac{I^{2}}{2\bar{\mu}B^{2}C^{2}}\right) = -\frac{A_{44}}{A}
- \frac{B_{44}}{B} - \frac{A_{4}B_{4}}{AB} - \Lambda
\end{equation}  
\begin{equation} 
\label{eq15} 
8\pi\left(\rho +  \frac{I^{2}}{2\bar{\mu}B^{2}C^{2}}\right) = \frac{A_{4}B_{4}}{AB}
+ \frac{A_{4}C_{4}}{AC} + \frac{B_{4}C_{4}}{BC} + \Lambda
\end{equation}  
The suffix $4$ by the symbols $A$, $B$ and $C$ denote differentiation with
respect to $t$.\\
%%%%%%%%%%%%%%%%%%%%%%%%%%%%%%%%%%%%%%%%%%%%%%%%%%%%%%%%%%%%%%%%%%%%%%%%%
%%%%%%%%%%%%%%%%%%%%%%%%%%  SECTION 3  %%%%%%%%%%%%%%%%%%%%%%%%%%%%%%%%%%
\section{Solution of the field equations}
Equations (\ref{eq12}) - (\ref{eq15}) represent a system of four equations
in six unknowns $A$, $B$, $C$, $p$, $\rho$ and $\Lambda$. To get a determinate 
solution, we need two extra conditions. First we assume
\begin{equation} 
\label{eq16} 
A = BC
\end{equation}  
From Eqs. (\ref{eq13}) and (\ref{eq14}), we get
\begin{equation} 
\label{eq17} 
\frac{B_{44}}{B} - \frac{C_{44}}{C} = \frac{A_{4}}{A}\left(\frac{C_{4}}{C} 
- \frac{B_{4}}{B}\right)
\end{equation}  
which leads to
\begin{equation} 
\label{eq18} 
\frac{\nu_{4}}{\nu} = \frac{\alpha}{\epsilon^{2}}
\end{equation}  
where $BC = \epsilon$, $\frac{B}{C} = \nu$, and $\alpha$ is a constant of integration.
From Eqs. (\ref{eq12}) and (\ref{eq13}), we get
\begin{equation} 
\label{eq19} 
\frac{K}{B^{2}C^{2}} = \frac{A_{44}}{A} - \frac{B_{44}}{B} + \frac{A_{4}C_{4}}{AC}
- \frac{B_{4}C_{4}}{BC}
\end{equation}  
where
\begin{equation} 
\label{eq20} 
K = -\frac{8\pi I^{2}}{\bar{\mu}}
\end{equation}  
By using the condition (\ref{eq16}) in (\ref{eq19}), we have
\begin{equation} 
\label{eq21} 
\frac{C_{44}}{C} + \frac{2B_{4}C_{4}}{BC} + \frac{C^{2}_{4}}{C^{2}} = 
\frac{K}{B^{2}C^{2}}
\end{equation}  
which leads to
\begin{equation} 
\label{eq22} 
\frac{\epsilon_{44}}{2\epsilon} +\frac{\epsilon^{2}_{4}}{2\epsilon^{2}} = 
\frac{K}{\epsilon^{2}}
\end{equation}  
Setting $\epsilon_{4}$ = $f(\epsilon)$ in Eq. (\ref{eq22}), we have
\begin{equation} 
\label{eq23} 
2\epsilon ff^{1} + 2f^{2} = 4K
\end{equation}  
which reduces to 
\begin{equation} 
\label{eq24} 
\frac{d}{d\epsilon}(f^{2}) + \frac{2f^{2}}{\epsilon} = \frac{4K}{\epsilon}
\end{equation}  
Integrating Eq. (\ref{eq24}) leads to
\begin{equation} 
\label{eq25} 
f^{2} = \frac{(2K\epsilon^{2} + N)}{\epsilon^{2}},
\end{equation}  
where $N$ is the constant of integration. Eq. (\ref{eq25}) on integration leads to 
\begin{equation} 
\label{eq26} 
\sqrt{(2K \epsilon^{2} + N)} = 2Kt + L,
\end{equation}  
where $L$ is the integrating constant. Eq. (\ref{eq26}) can be rewritten as
\begin{equation} 
\label{eq27} 
\epsilon = \sqrt{(at^{2} + bt + d)},
\end{equation}
where 
\begin{equation} 
\label{eq28} 
a = 2K;~ ~  b = 2L ~ ~ and ~ ~ d = \frac{(L^{2} - N)}{2K}.
\end{equation}   
From Eq. (\ref{eq18}), we get
\begin{equation} 
\label{eq29}
\frac{\nu_{4}}{\nu} = \frac{\alpha}{(at^{2} + bt + d)}
\end{equation}
Integrating Eq. (\ref{eq29}) leads to
\begin{equation} 
\label{eq30} 
\nu = \frac{1}{M} e^{\frac{2\alpha}{\sqrt{(4ad - b^{2})}} 
tan^{-1}\frac{(2at + b)}{\sqrt{(4ad - b^{2})}}}, ~ ~  if ~ ~ 4ad - b^{2} >0
\end{equation}  
where $M$ is a constant of integration.
\begin{equation} 
\label{eq31} 
\nu = L\left[\frac{(t + \frac{b}{2a}) - \frac{\sqrt{l}}{2a}}
{(t + \frac{b}{2a}) - \frac{\sqrt{l}}{2a}}\right]^{\frac{\alpha}{\sqrt{l}}}, 
~ ~ if ~ ~ 4ad - b^{2} < 0
\end{equation}  
where $l > 0$. 
\begin{equation} 
\label{eq32} 
\nu = e^{- \frac{\alpha}{a(t + \frac{b}{2a})}},~ ~  if ~ ~ 4ad - b^{2} = 0
\end{equation}  
%%%%%%%%%%%%%%%%%%%%%%%%%%%%%%%%%%%%%%%%%%%%%%%%%%%%%%%%%%%%%%%%%%%%%%%%%
%%%%%%%%%%%%%%%%%%%%%%%%%%%%  SUBSECTION 3.1   %%%%%%%%%%%%%%%%%%%%%%%%%%
\subsection{Case (i) : when~ $4ad - b^{2} > 0$.}
In this case, we have
\begin{equation} 
\label{eq33} 
A^{2} = at^{2} + bt + d
\end{equation}  
\begin{equation} 
\label{eq34} 
B^{2} = \frac{\sqrt{(at^{2} + bt + d)}}{M}e^{\frac{2\alpha}{\sqrt{(4ad - b^{2})}}
tan^{-1}\frac{(2at + b)}{\sqrt{(4ad - b^{2})}}}
\end{equation}  
\begin{equation} 
\label{eq35} 
C^{2} = \frac{M\sqrt{(at^{2} + bt + d)}}{\left[e^{\frac{2\alpha}{\sqrt{(4ad - b^{2})}}
tan^{-1}\frac{(2at + b)}{\sqrt{(4ad - b^{2})}}}\right]}
\end{equation}  
Hence, the geometry of the spacetime (\ref{eq1}) reduces to the form
\[
ds^{2} = - dt^{2} + (at^{2} + bt + d)dx^{2}\] 
\[
+ \frac{\sqrt{(at^{2} + bt + d)}}{M}
e^{\frac{2\alpha}{\sqrt{(4ad - b^{2})}}tan^{-1}\frac{(2at + b)}
{\sqrt{(4ad - b^{2})}}}dy^{2} \]
\begin{equation} 
\label{eq36}
+ \frac{M\sqrt{(at^{2} + bt + d)}}{\left[e^{\frac{2\alpha}{\sqrt{(4ad - b^{2})}}
tan^{-1}\frac{(2at + b)}{\sqrt{(4ad - b^{2})}}}\right]}dz^{2}
\end{equation}  
The effective pressure $\bar{p}$ and density $\rho$ for the model (\ref{eq36}) are
given by
\begin{equation} 
\label{eq37} 
8\pi(p - \xi \theta) = \frac{4(b^{2} - 4ad) + (2at + b)^{2} - 4 \alpha^{2}}
{16(at^{2} + bt + d)^{2}} - \frac{K}{2(at^{2} + bt + d)} - \Lambda
\end{equation}  
\begin{equation} 
\label{eq38} 
8\pi \rho =  \frac{5(2at + b)^{2} - 4 \alpha^{2}}
{16(at^{2} + bt + d)^{2}} + \frac{K}{2(at^{2} + bt + d)} + \Lambda
\end{equation}  
where $\theta$ is the scalar of expansion calculated for the flow vector $v^{i}$
and is given by
\begin{equation} 
\label{eq39} 
\theta = \frac{(2at + b)}{(at^{2} + bt + d)}
\end{equation}  
For the specification of $\xi$, now we assume that the fluid obeys an equation
of state of the form
\begin{equation} 
\label{eq40} 
p = \gamma \rho,
\end{equation} 
where $\gamma(0\leq\gamma \leq 1)$ is a constant.\\
Thus, given $\xi(t)$ we can solve the cosmological parameters. In most of the 
investigations involving bulk viscosity is assumed to be a simple power function of 
the energy density.\cite{ref32}$^-$\cite{ref34} 
\begin{equation}
\label{eq41}
\xi(t) = \xi_{0} \rho^{n},
\end{equation}
where $\xi_{0}$ and $n$ are constants. If $n = 1$, Eq. (\ref{eq41}) may correspond
to a radiative fluid.\cite{ref35} However, more realistic models\cite{ref36} are 
based on lying in the regime $0 \leq n \leq \frac{1}{2}$. On using Eq. (\ref{eq41}) 
in Eq. (\ref{eq37}), we obtain 
\begin{equation}
\label{eq42}
8\pi(p - \xi_{0}\rho^{n} \theta) = \frac{4(b^{2} - 4ad) + (2at + b)^{2} - 4 \alpha^{2}}
{16 T^{2}} - \frac{K}{2 T} - \Lambda
\end{equation}
where $T = at^{2} + bt + d$.\\
%%%%%%%%%%%%%%%%%%%%%%%%%%%%%%%%%%%%%%%%%%%%%%%%%%%%%%%%%%%%%%%%%%%%%%%%%%%%%%%%%%%%
%%%%%%%%%%%%%%%%%%%%%%% SUBSUBSECTION 3.1.1  %%%%%%%%%%%%%%%%%%%%%%%%%%%%%%%%%%%%%%%
\subsubsection{Model (i) :~ ~   $(\xi = \xi_{0})$.}
When $n$ = $0$, Eq. (\ref{eq41}) reduces to $\xi$ = $\xi_{0}$. 
With the use of Eqs. (\ref{eq38}), (\ref{eq39}) and (\ref{eq40}),
Eq. (\ref{eq42}) reduces to
\[
8\pi \gamma \rho = \frac{8\pi(2at + b)\xi_{0}}{T} 
+ \frac{4(b^{2} - 4ad) + (2at + b)^{2}}
{16T^{2}}\] 
\begin{equation}
\label{eq43}
- \frac{K}{2T} - \Lambda
\end{equation}
Eliminating $\rho(t)$ between Eqs. (\ref{eq38}) and (\ref{eq43}), we get
\[
(1 + \gamma)\Lambda = \frac{1}{16 T^{2}}\Big[(2at + b)^{2}(1 - 5\gamma) + 4(b^{2} - 4ad) 
+ 4\gamma\alpha^{2}\Big] \]
\begin{equation}
\label{eq44}
+ \frac{1}{2T}\Big[K(1 - \gamma) + 16\pi\xi_{0}(2at + b)\Big]
\end{equation}
%%%%%%%%%%%%%%%%%%%%%%%%%%%%%%%%%%%%%%%%%%%%%%%%%%%%%%%%%%%%%%%%%%%%%%%%%%%%%%%%%%%%%%
%%%%%%%%%%%%%%%%%%%%%%%   SUBSUBSECTION  3.1.2   %%%%%%%%%%%%%%%%%%%%%%%%%%%%%%%%%%%%
\subsubsection{Model (ii) :~ ~   $ (\xi = \xi_{0}\rho)$.}
When $n = 1$, Eq. (\ref{eq41}) reduces to $\xi = \xi_{0}\rho$. 
Eq. (\ref{eq42}), with the use of Eqs. (\ref{eq38}), (\ref{eq39}) and
(\ref{eq40}), reduces to
\[
8\pi \rho \left[\gamma - \frac{(2at + b)\xi_{0}}{T^{2}}\right] = 
\frac{4(b^{2} - 4ad) + (2at + b)^{2} - 4\alpha^{2}}{16T^{2}}\]
\begin{equation}
\label{eq45}
- \frac{K}{2T} - \Lambda
\end{equation}
Eliminating $\rho(t)$ between Eqs. (\ref{eq38}) and (\ref{eq45}), we get
\begin{eqnarray}
\Lambda &=& \frac{1}{16 \left[ \left( 1 + \gamma \right) T^2 - \left( 2at + b \right)
\xi_0 \right]}
\times \left[ 4 \left( b^2 - 4ad \right) + \left( 2at + b \right)^2 - \right. \nonumber \\ 
&&    4 \alpha^2 - 8KT  
  - \left( \gamma T^2 - \left( 2at + b \right) \xi_0 \right)   \nonumber \\ 
&& \left. \times \left\{ \frac{5 \left( 2at + b \right)^2 - 4 \alpha^2 + 8KT }{T^2} \right\} \right]
\label{eq46}
\end{eqnarray}
From Eqs. (\ref{eq44}) and (\ref{eq46}), we observe that the cosmological
constant is a decreasing function of time and it approaches a small value 
as time progresses (i. e., the present epoch), which explains the small
value of $\Lambda$ at present. In the model (i), we also find that $\Lambda$
is positive for all values of $\gamma$ lying in the interval $(0, \frac{1}{5})$
and $\alpha^{2} > 5(4ad - b^{2})$. In the model (ii), we see that the second
term of the Eq. (\ref{eq46}) decreases rapidly with expansion of the universe
and the first term will dominate over the second term. Thus, from Eq. (\ref{eq46})
it is also observed that $\Lambda$ is positive when $K < \frac{a}{2}$ and
$b^{2} > 8Kd + 4\alpha^{2}$. These small positive values of $\Lambda$ in both 
models (i) and (ii) are supported by results from recent supernovae observations 
(Perlmutter et al., [58] Riess et al., [59] Garnavich et al., [61] 
Schmidt et al. [64]).\\
\noindent {\bf Physical and Geometrical Features of the Models} \\
The component of the shear tensor $\sigma^{j}_{i}$ are given by
\[
\sigma^{1}_{1} = \frac{(2at + b)}{6T}
\]
\[
\sigma^{2}_{2} = \frac{(2at + b)}{12T} + \frac{\alpha}{2T}
\]
\[
\sigma^{3}_{3} = \frac{(2at + b)}{12T} - \frac{\alpha}{2T}
\]
\begin{equation}
\label{eq47}
\sigma^{4}_{4} = 0
\end{equation}
Thus
\begin{equation}
\label{eq48}
\sigma^{2} = \frac{(2at + b)^{2} - 12 \alpha^{2}}{48T}
\end{equation}
The rate of expansion $H_{i}$ in the direction of $x$, $y$, $z$ -axes  
are given by
\[
H_{1} = \frac{(2at + b)}{2T}
\]
\[
H_{2} = \frac{(2at + b)}{4T} + \frac{\alpha}{2T}
\]
\begin{equation}
\label{eq49}
H_{3} = \frac{(2at + b)}{4T} - \frac{\alpha}{2T}
\end{equation}
The model (\ref{eq36}) starts expanding at $t>-\frac{b}{2a}$ and the expansion
in the model decreases as time increases and stops at $t = \infty$. The model, 
in general, represents shearing and non-rotating universe. Since 
$\lim_{t\rightarrow \infty}\frac{\sigma}{\theta} \ne 0$, therefore, the model does 
not approach isotropy for large values of $t$. There is a singularity in the model
at $t$ = $0$ which is real physical singularity. \\
%%%%%%%%%%%%%%%%%%%%%%%%%%%%%%%%%%%%%%%%%%%%%%%%%%%%%%%%%%%%%%%%%%%%%%%%%%%%%%%%%%%%%
%%%%%%%%%%%%%%%%%%%%%%   SUBSECTION  3.2   %%%%%%%%%%%%%%%%%%%%%%%%%%%%%%%%%%%%%%%%
\subsection {Case (ii) : ~  when $4ad - b^{2} < 0 ~ i.e.~ 4ad - b^{2} = - l, ~l>0$.}
In this case the geometry of the spacetime (\ref{eq1}) takes the form
\[
ds^{2} = - dt^{2} + a\left[(t + \frac{b}{2a})^{2} - \frac{l}{4a^{2}}\right]dx^{2} \]
\[
+ a\left[(t + \frac{b}{2a})^{2} - \frac{l}{4a^{2}}\right]^{\frac{1}{2}}
\Big[\frac{(t + \frac{b}{2a}) - \frac{\sqrt{l}}{2a}}{(t + \frac{b}{2a}) + 
\frac{\sqrt{l}}{2a}}\Big]^{\frac{\alpha}{\sqrt{l}}} dy^{2} \]
\begin{equation}
\label{eq50}
+ a\left[(t + \frac{b}{2a})^{2} - \frac{l}{4a^{2}}\right]^{\frac{1}{2}}
\left[\frac{(t + \frac{b}{2a}) - \frac{\sqrt{l}}{2a}}{(t + \frac{b}{2a}) + 
\frac{\sqrt{l}}{2a}}\right]dz^{2}
\end{equation}
The effective pressure and density for the model (\ref{eq50}) are given by
\begin{equation}
\label{eq51}
8\pi(p - \xi \theta) = \frac{4a^{2}t^{2} + 5b^{2} - 4abt - 16ad}{16a^{2}T^{2}_{1}}
- \frac{\alpha^{2}}{16T_{1}} - \frac{K}{2\bar{\mu}aT_{1}} - \Lambda
\end{equation}
\begin{equation}
\label{eq52}
8\pi\rho = \frac{5(t + \frac{b}{2a})^{2}}{4T^{2}_{1}} - \frac{\alpha^{2}}{16T_{1}}
- \frac{K}{2\bar{\mu}aT_{1}} + \Lambda
\end{equation}
where the scalar expansion $\theta$ is given by
\begin{equation}
\label{eq53}
\theta = \frac{2(t + \frac{b}{2a})}{T_{1}}
\end{equation}
and 
\[
T_{1} = \left[(t + \frac{b}{2a})^{2} - \frac{l}{4a^{2}} \right]
\]
%%%%%%%%%%%%%%%%%%%%%%%%%%%%%%%%%%%%%%%%%%%%%%%%%%%%%%%%%%%%%%%%%%%%%%%%%%
%%%%%%%%%%%%%%%%%%%  SUBSUBSECTION  3.2.1  %%%%%%%%%%%%%%%%%%%%%%%%%%%%%%
\subsubsection{Model (i): ~ ~$(\xi = \xi_{0})$.}
When $n = 0$, Eq. (\ref{eq41}) reduces to $\xi = \xi_{0}$ and hence
Eq. (\ref{eq51}), with the use of Eqs. (\ref{eq40}), (\ref{eq52}) and 
(\ref{eq53}), leades to
\[
16\pi \gamma \rho = 16\pi\xi_{0}(t + \frac{b}{2a}) + \frac{1}{16a^{2}T^{2}_{1}}
\Big[4a^{2}t^{2}(1 + 5\gamma) + 5b^{2}(1 + \gamma) - 4at(1 - 5b\gamma) - 16ad\Big]\]
\begin{equation}
\label{eq54}
- \frac{\alpha^{2}(1 - \gamma)}{16T_{1}} -\frac{K(1 + \gamma)}{2\bar{\mu}aT_{1}}
-(1 - \gamma) \Lambda
\end{equation}
Eliminating $\rho(t)$ between Eqs. (\ref{eq52}) and (\ref{eq54}), we get
\[
(1 + \gamma)\Lambda = \frac{1}{16a^{2}T^{2}_{1}}\Big[4a^{2}t^{2}(1 - 5\gamma) 
+ 5b^{2}(1 -\gamma) - 4abt(1 + 5\gamma) - 16ad\Big]\]
\begin{equation}
\label{eq55} 
+\frac{\alpha^{2}(1 - \gamma)}{16T_{1}} + \frac{K(1 - \gamma)}{2\bar{\mu}aT_{1}}
- \frac{16\pi\xi_{0}(t + \frac{b}{2a})}{T_{1}}
\end{equation}
%%%%%%%%%%%%%%%%%%%%%%%%%%%%%%%%%%%%%%%%%%%%%%%%%%%%%%%%%%%%%%%%%%%%%%%%%%%%%%%%%%%
%%%%%%%%%%%%%%%%%%%%%%   SUBSUBSECTION  3.2.2  %%%%%%%%%%%%%%%%%%%%%%%%%%%%%%%%%%
\subsubsection{Model (ii) : ~ ~ ~ $(\xi = \xi_{0}\rho)$.}
When $n = 1$, Eq. (\ref{eq41}) reduces to $\xi = \xi_{0}\rho$. The Eq. (\ref{eq51}),
with the use of the Eqs. (\ref{eq40}), (\ref{eq52}) and (\ref{eq53}), reduces to
\[
8\pi \rho[a(1 + \gamma)T_{1} - (2at + b)\xi_{0}] = \frac{4a^{2}t^{2} + 5b^{2} - 
4abt - 16 ad}{16aT_{1}} +  \]
\begin{equation}
\label{eq56} 
\frac{5(2at + b)^{2}}{16aT_{1}} - \frac{a\alpha^{2}}{8} - \frac{K}{\bar{\mu}}
\end{equation}
Eliminating $\rho(t)$ between Eqs. (\ref{eq52}) and (\ref{eq54}), we get
\[
\Lambda[a(1 + \gamma)T_{1} - (2at + b)\xi_{0}] = \frac{4a^{2}t^{2} + 5b^{2} - 
4abt - 16 ad}{16aT_{1}} - \frac{a\alpha^{2}}{16} - \frac{K}{2\bar{\mu}} \]
\begin{equation}
\label{eq57} 
- {a\gamma T_{1} - (2at + b)\xi_{0}}\left[\frac{5(t + \frac{b}{2a})^{2}}{4T^{2}_{1}}
 - \frac{\alpha^{2}}{16T^{2}_{1}} - \frac{K}{2\bar{\mu}aT_{1}}\right]
\end{equation}
In both models (i) and (ii), we have observed from Eqs. (\ref{eq55}) and 
(\ref{eq57}) that the cosmological constant is a decreasing function of time 
and it approaches a small value as time progresses (i.e., the present epoch), 
which explains the small value of $\Lambda$ at present.\\
\noindent {\bf The Physical and Geometrical Features of the Models}\\
The non-vanishing components of shear tensor $(\sigma^{j}_{i})$ are given by
\begin{equation}
\label{eq58} 
\sigma^{1}_{1} = \frac{2(t + \frac{b}{2a})}{(t + \frac{b}{2a})^{2} - \frac{l}{4a^{2}}}
\end{equation}
\begin{equation}
\label{eq59} 
\sigma^{2}_{2} = \frac{1}{12}\left[\frac{3\alpha}{(t + \frac{b}{2a})^{2} - \frac{l}
{4a^{2}}} - \frac{2(t + \frac{b}{2a})}{(t + \frac{b}{2a})^{2} - \frac{l}{4a^{2}}}\right]
\end{equation}
\begin{equation}
\label{eq60} 
\sigma^{3}_{3} = - \frac{1}{12}\left[\frac{3\alpha}{(t + \frac{b}{2a})^{2} - \frac{l}
{4a^{2}}} + \frac{2(t + \frac{b}{2a})}{(t + \frac{b}{2a})^{2} - \frac{l}{4a^{2}}}\right]
\end{equation}
The models start expanding with a big bang at $t = -\frac{b}{2a} + \frac{\sqrt{l}}{2a}$
 and the expansion in the models decreases with time. The expansion in the models 
stops at $t = \infty$. Since $lim_{t \rightarrow \infty} {\frac{\sigma}{\theta}} \ne 0$.
Therefore, the models do not approach isotropy for large values of $t$. \\
%%%%%%%%%%%%%%%%%%%%%%%%%%%%%%%%%%%%%%%%%%%%%%%%%%%%%%%%%%%%%%%%%%%%%%%%%
%%%%%%%%%%%%%%%%%%%%%%%%%%%%%%  SECTION 3.3 %%%%%%%%%%%%%%%%%%%%%%%%%%%%%%%
\subsection {Case (iii) : ~  when $4ad - b^{2} = 0$.}
In this case the metric (\ref{eq1}) reduces to
\begin{equation}
\label{eq61} 
ds^{2} = - dT^{2}_{2} + T^{2}_{2}~dX^{2} + T_{2}~ e^{-\frac{\alpha}{aT_{2}}}~dY^{2}
+ T_{2}~ e^{\frac{\alpha}{aT_{2}}}~dZ^{2},
\end{equation}
where
\[
T_{2} = t + \frac{b}{2a}
\]
The effective pressure $\bar{p}$ and density $\rho$ for the model (\ref{eq61})
are given by
\begin{equation}
\label{eq62} 
8\pi \bar{p} = 8\pi(p - \xi \theta) = \frac{1}{4T^{2}_{2}} - 
\frac{\alpha^{2}}{4a^{2}T^{4}_{2}} - \frac{K}{2aT^{2}_{2}}
- \Lambda,
\end{equation}
\begin{equation}
\label{eq63} 
8\pi \rho  = \frac{5}{4T^{2}_{2}} - 
\frac{\alpha^{2}}{4a^{2}T^{4}_{2}} + \frac{K}{2aT^{2}_{2}}
+ \Lambda,
\end{equation}
where the scalar expansion $\theta$ is given by
\begin{equation}
\label{eq64} 
\theta = \frac{2}{T_{2}}
\end{equation}
%%%%%%%%%%%%%%%%%%%%%%%%%%%%%%%%%%%%%%%%%%%%%%%%%%%%%%%%%%%%%%%%%%%%%%%%%
%%%%%%%%%%%%%%%%%%%% SUBSUBSECTION 3.3.1    %%%%%%%%%%%%%%%%%%%%%%%%%%%%%%%%%%
\subsubsection {Model (i): ~ ~$(\xi = \xi_{0})$.}
When $n = 0$, Eq. (\ref{eq41}) gives $\xi$ = $\xi_{0}$ and hence
Eq. (\ref{eq62}), with the use of Eqs. (\ref{eq40}), (\ref{eq63}) and (\ref{eq64}),
reduces to
\begin{equation}
\label{eq65} 
\rho = \frac{1}{8\pi(1 + \gamma)T_{2}}\left[16\pi \xi_{0} + \frac{3}{2T_{2}}
- \frac{\alpha^{2}}{2a^{2}T^{3}_{2}}\right]
\end{equation}
Eliminating $\rho(t)$ between Eqs. (\ref{eq63}) and (\ref{eq65}), we get
\begin{equation}
\label{eq66} 
(1 + \gamma)\Lambda = \frac{16\pi \xi_{0}}{T_{2}} + 
\frac{1}{4aT^{2}_{2}}\left[a(1 - 5\gamma ) - 2(1 + \gamma)K\right] 
- \frac{(1 - \gamma)\alpha^{2}}{4a^{2}T^{4}_{2}}
\end{equation}
%%%%%%%%%%%%%%%%%%%%%%%%%%%%%%%%%%%%%%%%%%%%%%%%%%%%%%%%%%%%%%%%%%%%%%%%%
%%%%%%%%%%%%%%%%%%%%  SUBSUBSECTION 3.3.2    %%%%%%%%%%%%%%%%%%%%%%%%%%%%%%%%%%%
\subsubsection {Model (ii) : ~ ~ ~ $(\xi = \xi_{0}\rho)$.}
When $n = 1$, Eq. (\ref{eq41}) gives $\xi$ = $\xi_{0} \rho$ and hence
Eq. (\ref{eq62}), with the use of Eqs. (\ref{eq40}), (\ref{eq63}) and (\ref{eq64}),
reduces to
\begin{equation}
\label{eq67} 
\rho = \frac{1}{8\pi[(1 + \gamma)T_{2} - 2 \xi_{0}]} \left[\frac{3}{2T_{2}}
- \frac{\alpha^{2}}{2a^{2}T^{3}_{2}}\right]
\end{equation}
Eliminating $\rho(t)$ between Eqs. (\ref{eq63}) and (\ref{eq67}), we get
\begin{eqnarray}
\label{eq68}
\Lambda &= & \frac{1}{\left\{\left( 1 + \gamma \right) T_2 - 2\xi_0\right\}T_2} 
\left[\frac{\left( 1 - 5\gamma\right) }{4}\right.
- \frac{\left(1 + \gamma \right)K}{2a} - \frac{\left(1 - \gamma \right)
\alpha^2}{4 a^2 T^2_2} + \nonumber \\
&&\left. \frac{2\xi_0}{T_2}\left\{\frac{5}{4} + \frac{K}{2 a} - 
\frac{\alpha^2}{4a^2 T^2_2}\right\}\right]
\end{eqnarray}
From Eqs. (\ref{eq66}) and (\ref{eq68}), we observe that the cosmological
constant is a decreasing function of time and it approaches a small value 
as time progresses (i. e., the present epoch), which explains the small
value of $\Lambda$ at present. In the model (i), from Eq. (\ref{eq66}), 
it can be seen that first and second terms dominate over the third term. 
Further we also find that $\Lambda$ is positive for all values of $\gamma$ 
lying in the interval $(0, \frac{1}{5})$ and $ K < \frac{a(1 - 5 \gamma)}
{2(1 + \gamma)}$. In the model (ii), from Eq. (\ref{eq68}), one can easily
see that that for all value of $\gamma$ lying in the interval $(0, \frac{1}{5})$
 and $\xi_{0} < \frac{(1 + \gamma)b}{4a}$, $ K < \frac{a(1 - 5 \gamma)}
{2(1 + \gamma)}$, $\Lambda$ is positive. These small positive values of 
$\Lambda$ are consistent by the results from  recent supernovae observations. \\
\noindent {\bf The Physical and Geometrical Features of the Models} \\
The non-vanishing components of shear tensor $(\sigma^{j}_{i})$ are given by
\begin{equation}
\label{eq69}
\sigma^{1}_{1} = \frac{1}{3T_{2}}
\end{equation}
\begin{equation}
\label{eq70}
\sigma^{2}_{2} = \frac{1}{6}\left[\frac{3\alpha}{aT^{2}_{2}} - \frac{1}{T_{2}}\right]
\end{equation}
\begin{equation}
\label{eq71}
\sigma^{2}_{2} = - \frac{1}{6}\left[\frac{3\alpha}{aT^{2}_{2}} + \frac{1}{T_{2}}\right]
\end{equation}
The models start expanding with a big bang $t$ = $-\frac{b}{2a}$ and the expansion
in the models decreases with time increase. The expansion in the models stop at
$t$ = $\infty$. Since $lim_{t \rightarrow \infty} {\frac{\sigma}{\theta}} \ne 0$.
Therefore, the models do not approach isotropy for large values of $t$.\\
%%%%%%%%%%%%%%%%%%%%%%%%%%%%%%%%%%%%%%%%%%%%%%%%%%%%%%%%%%%%%%%%%%%%%%%%%%%%
%%%%%%%%%%%%%%%%%%%%%%%%%%%%%%%%%% SECTION 4 %%%%%%%%%%%%%%%%%%%%%%%%%%%%%%%
\section{Solution of the field equations in absence of magnetic field}
In the absence of magnetic field, the metric (\ref{eq36}) reduces to
\begin{equation}
\label{eq72}
ds^{2} = - dt^{2} + (bt + \beta)dx^{2} + \frac{l\sqrt{(bt + \beta)}}{M}dy^{2}
+ \frac{M\sqrt{(bt + \beta)}}{l}dz^{2},
\end{equation}
where
\[
l = e^{\frac{2\alpha}{\sqrt{h - b^{2}}}tan^{-1}\frac{b}{\sqrt{h - b^{2}}}}; 
~ ~ ~ a = 2K \]
\begin{equation}
\label{eq73}
d = \frac{L^{2} - N}{2K} = \beta ~\frac{\sin  2K}{2K}; ~ ~ ~ 4ad = 4(L^{2} - N) = h
\end{equation}
Thus when $K \rightarrow 0$ then $d \rightarrow \beta$. In absence of the
magnetic field i.e. when $K \rightarrow 0$ then effective pressure $\bar p$ 
and density $\rho$ for the model (\ref{eq72}) are given by 
\begin{equation}
\label{eq74}
8\pi \bar{p} = 8\pi (p - \xi \theta) = \frac{b^{2} - d}{4(bt + \beta)^{2}} 
+ \frac{b^{2}}{16(bt + \beta)^{2}} - \frac{\alpha^{2}}{4(bt + \beta)^{2}} 
- \Lambda,
\end{equation} 
\begin{equation}
\label{eq75}
8\pi\rho = \frac{5b^{2}}{16(bt + \beta)^{2}} - \frac{\alpha^{2}}{4(bt + \beta)^{2}}
+ \Lambda, 
\end{equation}
where scalar expansion $\theta$ is given by
\begin{equation}
\label{eq76}
\theta = \frac{b}{bt + \beta}
\end{equation}
%%%%%%%%%%%%%%%%%%%%%%%%%%%%%%%%%%%%%%%%%%%%%%%%%%%%%%%%%%%%%%%%%%%%%%%%%%%%%%%%%%
%%%%%%%%%%%%%%%%%%%%%%%%%%%%%%% SUBSECTION 4.1  %%%%%%%%%%%%%%%%%%%%%%%%%%%%%%%%%
\subsection{Model (i): ~ ~$(\xi = \xi_{0})$.}
When $n = 0$, Eq. (\ref{eq41}) gives $\xi$ = $\xi_{0}$  and hence
Eq. (\ref{eq74}), with the use of Eqs. (\ref{eq40}), (\ref{eq75}) and (\ref{eq76}),
reduces to
\begin{equation}
\label{eq77} 
8\pi (1 + \gamma)\rho = \frac{8\pi b \xi_{0}}{(bt + \beta)} + 
\frac{5b^{2} - 4\alpha^{2} -2d}{8(bt + \beta)^{2}}
\end{equation}
Eliminating $\rho(t)$ between Eqs. (\ref{eq75}) and (\ref{eq77}), we get
\begin{equation}
\label{eq78} 
(1 + \gamma)\Lambda = \frac{8\pi b \xi_{0}}{(bt + \beta)} - 
\frac{d}{4(bt + \beta)^{2}}
\end{equation}
%%%%%%%%%%%%%%%%%%%%%%%%%%%%%%%%%%%%%%%%%%%%%%%%%%%%%%%%%%%%%%%%%%%%%%%%%%%%%%%%%%%%
%%%%%%%%%%%%%%%%%%%%%%%%%%  SUBSECTION  4.2  %%%%%%%%%%%%%%%%%%%%%%%%%%%%%%%%%%%%
\subsection{Model (ii) : ~ ~ ~ $(\xi = \xi_{0}\rho)$.}
When $n = 1$, Eq. (\ref{eq41}) reduces to $\xi = \xi_{0}\rho$. 
Eq. (\ref{eq74}), with the use of Eqs. (\ref{eq40}), (\ref{eq75}), and 
(\ref{eq76}), reduces to
\begin{equation}
\label{eq79} 
8\pi \rho = \frac{(5b^{2} - 4\alpha^{2} -2d)}{8(bt + \beta)[(1 + \gamma)(bt + \beta) 
- b \xi_{0}]}
\end{equation}
Eliminating $\rho(t)$ between Eqs. (\ref{eq75}) and (\ref{eq79}), we get
\begin{equation}
\label{eq80} 
\Lambda = \frac{\left[(1 - \gamma)(5b^{2} - 4\alpha^{2}) -4ad + 
\frac{(5b^{2} - 4\alpha^{2})b\xi_{0}}{(bt + \beta)}\right]}{16(bt + \beta)
\left[(1 + \gamma)(bt + \beta) - b \xi_{0}\right]}
\end{equation}
In both the models (i) and (ii), from Eqs. (\ref{eq78}) and (\ref{eq80}), we 
observe that the cosmological constant is a decreasing function of time and 
it approaches a small value as time progresses (i. e., the present epoch), 
which explains the small value of $\Lambda$ at present. In the model (i), 
from Eq. (\ref{eq78}), it can be seen that for all time $t > \frac{d}{32\pi b^{2}
 \xi_{0}} - \frac{\beta}{B}$, the cosmological ``constant'' is positive definite. 
In model (ii), Eq. (\ref{eq80}) suggest that the positivity of $\Lambda$
demands the relation among constant as $ b^{2} > \frac{4}{5} (\alpha^{2} + ad)$
 and $\xi_{0} < \frac{(1 + \gamma)\beta}{b}$. These small positive values
of $\Lambda$ in both models are supported by the results from recent supernovae 
observations.\\ 
\noindent{\bf The Physical and Geometrical Features of the Models.}\\
In the absence of magnetic field, the shear $(\sigma^{j}_{i})$ are given by
\begin{equation}
\label{eq81} 
\sigma^{2} = \frac{b^{2} + 12 \alpha^{2}}{48(bt + \beta)^{2}}
\end{equation}
Here, $lim_{t \rightarrow \infty} {\frac{\sigma}{\theta}} \ne 0$. The model 
does not approach isotropy for large values of $t$ in absence of magnetic 
field also. \\
%%%%%%%%%%%%%%%%%%%%%%%%%%%%%%%%%%%%%%%%%%%%%%%%%%%%%%%%%%%%%%%%%%%%%%%%%%%%%%%%%%%%
%%%%%%%%%%%%%%%%%%%%%  SECTION 5   %%%%%%%%%%%%%%%%%%%%%%%%%%%%%%%%%%%%%%%%%%%%%%%%%
\section{Conclusions} 
We have obtained a new class of Bianchi type I anisotropic magnetized cosmological
models with a bulk viscous fluid as the source of matter. Generally, the models are
expanding, shearing and non-rotating. In all these models, we observe that they do 
not approach isotropy for large values of time $t$ either in the presence or in the 
absence of magnetic field. \\
The cosmological constant in all models given in Sec. 3 and Sec. 4 are decreasing 
function of time and they all approach a small value as time increases (i.e., the
present epoch). The values of cosmological ``constant'' for these models are
found to be small and positive which are supported by the results from recent
supernovae observations recently obtained by the High - Z Supernova Team and 
Supernova Cosmological Project ( Garnavich {\it et al.};\cite{ref61} Perlmutter 
{\it et al.};\cite{ref58} Riess {\it et al.};\cite{ref59} Schmidt {\it et al.}
\cite{ref64}). Thus, with our approach, we obtain a physically relevant decay 
law for the cosmological constant unlike other investigators where {\it adhoc} 
laws were used to arrive at a mathematical expressions for the decaying vacuum 
energy. Thus our models are more general than those studied earlier.  
\nonumsection{Acknowledgements} 
\noindent One of the authors (A. Pradhan) thanks to the Inter-University Centre 
for Astronomy and Astrophysics, India for providing  facility under
Associateship Programme where this work was carried out. We would like
to thank G. P. Singh for helpful discussions and also to Christos G. Tsagas
for pointing a fruitful comment.  \\
\newline
\newline
\nonumsection{References}

\end{document}